\documentclass[aps,prd,preprintnumbers,showpacs,superscriptaddress,nofootinbib,amsmath,amssymb,floats,floatfix,showkeys,notitlepage,longbibliography]{revtex4-1}
\usepackage{comment}
\usepackage{graphicx}
\usepackage{subfigure}
\usepackage{palatino}
\usepackage[commandnameprefix=always]{changes}
\usepackage{hyperref}
\hypersetup{colorlinks=true,linkcolor=red,urlcolor=red,citecolor=red}
\usepackage[toc,page]{appendix}
\usepackage[normalem]{ulem}

\usepackage{lipsum}
\usepackage{graphicx}
\usepackage{subfigure}
\usepackage{palatino}
\usepackage{float}
\usepackage{sans}
\usepackage{adjustbox}
\usepackage{latexsym}
\usepackage{amsmath}
\usepackage{amssymb}
\usepackage{amsfonts}
\usepackage{dcolumn}
\usepackage{bm}
\usepackage{tikz}
\usepackage{bigints}
\usepackage{array,tabularx,multirow,booktabs}
\usepackage[tracking=true]{microtype}
\SetTracking{}{500}
\SetTracking{encoding={*}, shape=sc}{40}
\allowdisplaybreaks
\usepackage{adjustbox}
\usepackage{latexsym}
\usepackage{amsmath}
\usepackage{amssymb}
\usepackage{amsfonts}
\usepackage{dcolumn}
\usepackage{bm}
\usepackage{tikz}
\usepackage{bigints}
\usepackage{array,tabularx,multirow,booktabs}
\usepackage[tracking=true]{microtype}
\usepackage{color}
\allowdisplaybreaks

\begin{document}

\title{Topology of Holographic Thermodynamics within Non-extensive Entropy}

\author{Saeed Noori Gashti
}
\email{saeed.noorigashti@stu.umz.ac.ir}
\affiliation{School of Physics, Damghan University, P. O. Box 3671641167, Damghan, Iran}
\affiliation{Department of Physics, Faculty of Basic
Sciences, University of Mazandaran\\ P. O. Box 47416-95447, Babolsar, Iran;\\ E-mail: saeed.noorigashti@stu.umz.ac.ir}

\begin{abstract}
In this paper, we delve into the thermodynamic topology of AdS Reissner-Nordstr$\ddot{o}$m (R-N) black holes by employing nonextensive entropy frameworks, specifically R$\acute{e}$nyi (with nonextensive parameter $\lambda$) and Sharma-Mittal entropy (with nonextensive parameter $\alpha, \beta$).
Our investigation spans two frameworks: bulk boundary and restricted phase space (RPS) thermodynamics. In the bulk boundary framework,  we face singular zero points revealing topological charges influenced by the free parameter $(\lambda)$ with a positive topological charge $(\omega = +1)$ and the total topological charge $(W = +1)$, indicating the presence of a single stable on-shell black hole. Further analysis shows that when $(\lambda)$ is set to zero, the equations align with the Bekenstein-Hawking entropy structure, demonstrating different behaviors with multiple topological charges $(\omega = +1, -1, +1)$. Notably, increasing the parameter $\alpha$ in Sharma-Mittal entropy results in multiple topological charges $(\omega = +1, -1, +1)$ with the total topological charge $(W = +1)$. Conversely, increasing $(\beta)$ reduces the number of topological charges, maintaining the total topological charge $(W = +1)$. Extending our study to the restricted phase space, we observe consistent topological charges $(\omega = +1)$ across all conditions and parameters. This consistency persists even when reducing to Bekenstein-Hawking entropy, suggesting similar behaviors in both non-extended and Hawking entropy states within RPS.
\end{abstract}

\date{\today}



\maketitle
%\tableofcontents
\section{Introduction}
Wei et al. \cite{a19,a20}  recently introduced a novel technique for exploring the topological charge of black holes. This method conceptualizes black hole solutions as topological defects within the thermodynamic parameter space. By leveraging the generalized off-shell free energy, they were able to categorize black holes based on their topological charge, which is determined by the winding numbers of these defects. Black holes with positive winding numbers are locally stable, whereas those with negative winding numbers are locally unstable \cite{a19,a20}. This new approach offers a unique viewpoint on the thermodynamic stability of black holes and provides valuable insights into the phase transitions and critical phenomena in black hole thermodynamics. It has been applied to various black holes, including those in anti-de Sitter (AdS) spacetime, uncovering new types of critical points and phase behaviors. The topological method for black hole thermodynamics has become popular due to its straightforwardness in examining thermodynamic properties. It has been utilized to investigate the Hawking-Page phase transition of Schwarzschild-AdS black holes and their holographic counterparts, which relate to the confinement-deconfinement transition in gauge theories. Quantum gravity corrections, expressed through higher-derivative terms, have been studied for black holes in Einstein-Gauss-Bonnet and Lovelock gravity. These corrections shed light on the behavior of black holes in higher-dimensional spacetimes and the effects of quantum gravity. Although these studies mainly focus on static black holes, the topological approach has also been extended to rotating black holes, offering significant insights into their thermodynamic properties and stability \cite{20a,21a,22a,23,24,25,26,27}. In this article, we aim to explore the topology of holographic thermodynamics using non-extensive entropies such as R$\acute{e}$nyi and Sharma-Mittal entropy. Our objective is to identify the topological class of these black holes and compare it with the Bekenstein-Hawking entropy. Non-extensive entropy, often linked with Tsallis entropy, is a generalization of the traditional Boltzmann-Gibbs entropy. This concept was introduced by Constantino Tsallis in 1988 to address systems where the conventional assumptions of extensive entropy do not apply. In classical thermodynamics, entropy is extensive, meaning it scales linearly with the system's size. However, many physical systems exhibit non-extensive behavior due to long-range interactions, fractal structures, or other complexities \cite{aa,bb,cc}. Non-extensive entropy has been applied to various astrophysical phenomena, including the distribution of stellar objects and the dynamics of galaxy clusters. It aids in modeling systems where gravitational interactions are long-range and cannot be described by extensive entropy. Non-extensive entropy extends information theory concepts to systems with non-standard probability distributions. It is utilized in coding theory, data compression, and the analysis of complex networks \cite{aa,bb,cc}. Holographic thermodynamics is a framework that applies the principles of holography to the study of black hole thermodynamics. This approach often involves the AdS/CFT correspondence, which posits a relationship between a gravitational theory in an anti-de Sitter (AdS) space and a conformal field theory (CFT) on its boundary. This duality allows physicists to study complex gravitational systems using quantum field theories' simpler, well-understood properties. So, one can study two spaces with some features viz bulk-boundary, and restricted phase space. Bulk-boundary correspondence is a principle that connects the properties of a bulk system (like a black hole in AdS space) with those of its boundary (the CFT). This correspondence is crucial in understanding topological phases of matter and has applications in condensed matter physics and high-energy physics. It essentially states that the behavior of a system's boundary can reveal information about the bulk properties. Restricted phase space thermodynamics is a newer formalism that modifies traditional black hole thermodynamics by fixing certain parameters, such as the AdS radius, as constants. This approach eliminates the need for pressure and volume as thermodynamic variables, instead using the central charge and chemical potential. This formalism maintains the Euler relation equation, providing a consistent framework for studying black hole thermodynamics \cite{709,710,711,712,713,714,715,716,717}. Based on these explanations, we will organize the article as follows; Section 2 will introduce the concepts of non-extensive entropies, specifically R$\acute{e}$nyi and Sharma-Mittal entropies. In Section 3, we will explain the thermodynamic topology using the generalized Helmholtz free energy method. In Section 4, we will provide a concise overview of the black hole model within the contexts of bulk-boundary correspondence and restricted phase space. So, this section involve the calculation and discussion of the thermodynamic topography of the model, focusing on non-extensive entropies. Finally, Section 5 will present our conclusions and summarize the key findings.
\section{Non-extensive Entropy}
Non-extensive entropy, introduced by Constantino Tsallis, is an extension of the traditional Boltzmann-Gibbs entropy. This concept is particularly useful for systems that exhibit non-linearity and a strong dependence on initial conditions. Unlike Boltzmann-Gibbs entropy, which assumes extensive properties (i.e., entropy scales linearly with system size), non-extensive entropy can handle systems where this linearity does not hold. It has applications in various fields, including theoretical physics, cosmology, and statistical mechanics, and is especially relevant for systems with long-range interactions, fractal structures, or memory effects \cite{a33}.
R$\acute{e}$nyi entropy, denoted as $S_R$, is defined as \cite{a25},
\begin{equation}\label{N1}
S_R = \frac{1}{\lambda} \ln(1 + \lambda S_{BH})
\end{equation}
where $\lambda$ is the non-extensive parameter. The valid range for $\lambda$ is $ -\infty < \lambda < 1$, as values outside this range render the entropy function ill-defined due to its convex nature. In the context of black hole thermodynamics using R$\acute{e}$nyi statistics, the entropy $ S_R $ is well-defined for $ 0 < \lambda < 1$. Within this range, $\lambda$ exhibits favorable thermodynamic properties, as highlighted in recent research. When the R$\acute{e}$nyi parameter $\lambda$ approaches zero, the generalized off-shell free energy converges to the Boltzmann-Gibbs statistics case \cite{a26,a27,aa27}. Sharma-Mittal entropy (SM), which generalizes both R$\acute{e}$nyi and Tsallis entropies, offers intriguing insights in cosmological studies. This generalization is particularly useful for describing the current accelerated expansion of the universe by effectively utilizing vacuum energy. Although non-extensive entropies have been used to study the thermodynamic properties of black holes, Sharma-Mittal entropy has not yet been applied in this context. This gap motivates us to explore the thermodynamic properties of black holes, considered as strongly coupled gravitational systems, using SM entropy. The SM entropy is defined as \cite{a28,a29,a30},
\begin{equation}\label{N2}
S_{SM} = \frac{1}{\alpha} \left( (1 + \beta S_T)^\frac{\alpha}{\beta} - 1 \right)
\end{equation}
In this expression, $S_T = \frac{A}{4}$, where $A = 4\pi r^2 $ and $r$ is the horizon radius, represents the Tsallis entropy. The parameters $\alpha$ and $\beta$ are free and must be determined by fitting the results with observational data. Notably, when $\alpha \to 0$ and $\alpha \to \beta$, Sharma-Mittal entropy reduces to R$\acute{e}$nyi and Tsallis entropies, respectively.
\section{F-Method}
Recent developments have introduced novel techniques for examining critical points and phase transitions in black hole thermodynamics. One significant approach is the topological method, particularly Duan's topological current $\phi$-mapping theory, which offers a topological perspective in thermodynamics. Wei et al. have pioneered methods for studying topological thermodynamics, emphasizing generalized free energy functions. Given the relationship between mass and energy in black holes, the generalized free energy function is expressed as \cite{a19,a20},
\begin{equation}\label{F1}
\mathcal{F} = M - \frac{S}{\tau},
\end{equation}
where $\tau$ denotes the Euclidean time period, and $T$, the inverse of $\tau$, represents the temperature of the ensemble. The generalized free energy is on-shell only when $\tau = \tau_{H} = \frac{1}{T_{H}}$. A vector $\phi$ is constructed as follows,
\begin{equation}\label{F2}
\phi = \left(\frac{\partial \mathcal{F}}{\partial r_{H}}, -\cot \Theta \csc \Theta \right).
\end{equation}
In this context, $\phi^{\Theta}$ diverges, and the vector direction points outward at $\Theta = 0$ and $\Theta = \pi$. The ranges for $r_{H}$ and $\Theta$ are $0 \leq r_{H} \leq \infty$ and $0 \leq \Theta \leq \pi$, respectively. Using Duan's $\phi$-mapping topological current theory, a topological current can be defined as:
\begin{equation}\label{F3}
j^{\mu} = \frac{1}{2\pi} \varepsilon^{\mu\nu\rho} \varepsilon_{ab} \partial_{\nu} n^{a} \partial_{\rho} n^{b}, \quad \mu, \nu, \rho = 0, 1, 2,
\end{equation}
where $n = (n^1, n^2)$, and $n^1 = \frac{\phi^r}{|\phi|}$, $n^2 = \frac{\phi^\Theta}{|\phi|}$. From the conservation equation, we observe that $j^{\mu}$ is non-zero only when $\phi = 0$. After some calculations, the topological number or total charge $W$ can be expressed as,
\begin{equation}\label{F4}
W = \int_{\Sigma} j^{0} d^2 x = \sum_{i=1}^{n} \beta_{i} \eta_{i} = \sum_{i=1}^{n} \omega_{i}.
\end{equation}
Here, $\beta_i$ denotes the positive Hopf index, which counts the loops of the vector $\phi^a$ in the $\phi$-space when $x^\mu$ is near the zero point $z_i$. Meanwhile, $\eta_i = sign(j^0(\phi/x)_{z_i}) = \pm 1$. The quantity $\omega_i$ represents the winding number for the $i$-th zero point of $\phi$ in $\Sigma$.
\section{AdS Reissner-Nordstr$\ddot{o}$m black holes}
AdS Reissner-Nordstr$\ddot{o}$m (AdS R-N) black holes are distinguished by their presence in AdS space and their Abelian electric charge. These black holes are solutions to the Einstein-Maxwell equations with a negative cosmological constant. The metric for an AdS R-N black hole is given by,
\begin{equation}\label{M1}
ds^2 = -f(r) dt^2 + \frac{dr^2}{f(r)} + r^2 d\Omega^2
\end{equation}
%----------------------------------------------------------------------
\subsection{Bulk boundary thermodynamics}
In the framework of bulk boundary, we will have
\begin{equation}\label{M2}
f(r) = 1 - \frac{2GM}{r} + \frac{Gq^2}{r^2} + \frac{r^2}{l^2}
\end{equation}
The entropy $S$ and the radius $l$ of AdS space are defined as?
\begin{equation}\label{M3}
S = \frac{r_h^2 \pi}{G}, \quad l = \frac{1}{4}\sqrt{\frac{6}{P G \pi}}.
\end{equation}
The Hawking temperature $ T $ for an AdS R-N black hole is,
\begin{equation}\label{M5}
T = \frac{2\big( - \frac{3q^2}{8P \pi} + \frac{3r_h^2}{8P G \pi} + 3r_h^4\big)P G}{3r_h^3}
\end{equation}
In this context, $G$ is variable and defined as:
\begin{equation}\label{M6}
G = \frac{r_h^2}{8P \pi r_h^4 + 3q^2}
\end{equation}
The mass $M$ and Helmholtz free energy for this black hole are,
\begin{equation}\label{M7}
M = \frac{4 \left( \frac{3q^2}{8P \pi} + \frac{3r_h^2}{8P G \pi} + r_h^4 \right)P\pi}{3r_h}
\end{equation}
Here, $M$ is the mass, $q$ is the charge, and $l$ is the AdS radius. This metric describes the spacetime geometry around the black hole, which has a singularity at $r = 0$ and one or two horizons depending on the values of $M$, $q$, and $l$. The horizons are the surfaces where $f(r_h) = 0 $, with the outermost horizon being the event horizon, which determines the black hole's size.
\subsubsection{Thermodynamic topology within R$\acute{e}$nyi statistics}
\begin{figure}[h!]
 \begin{center}
 \subfigure[]{
 \includegraphics[height=3cm,width=2.8cm]{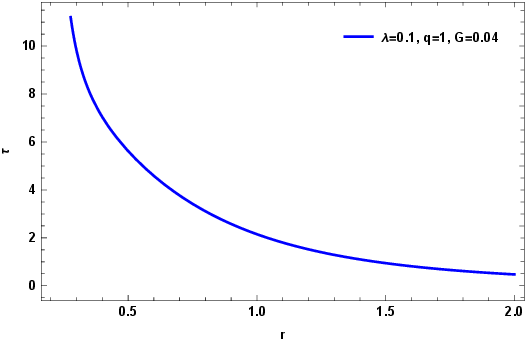}
 \label{1a}}
 \subfigure[]{
 \includegraphics[height=3cm,width=2.5cm]{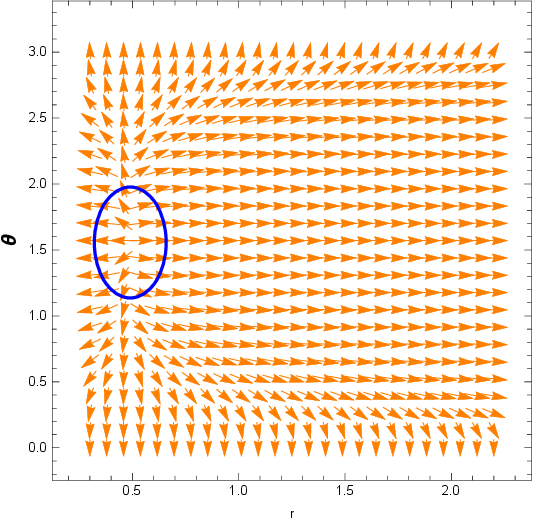}
 \label{1b}}
 \subfigure[]{
 \includegraphics[height=3cm,width=2.5cm]{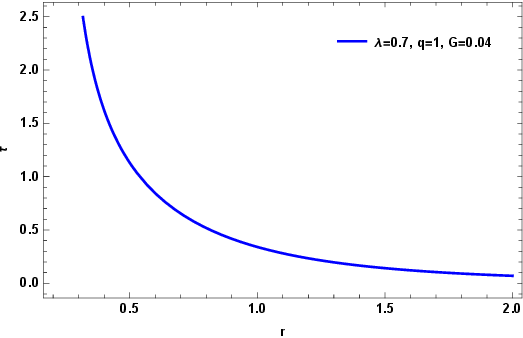}
 \label{1c}}
 \subfigure[]{
 \includegraphics[height=3cm,width=2.5cm]{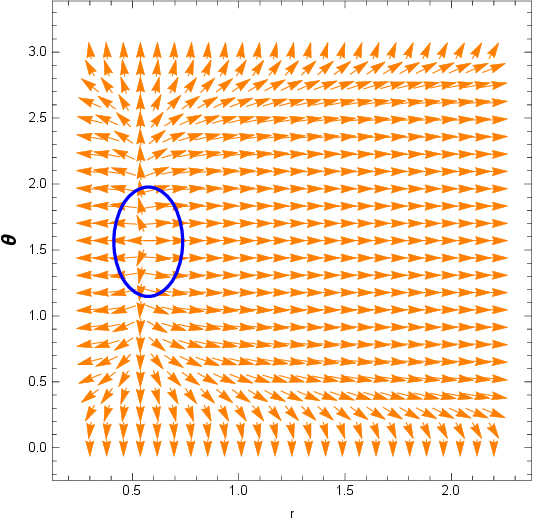}
 \label{1d}}
 \subfigure[]{
 \includegraphics[height=3cm,width=2.5cm]{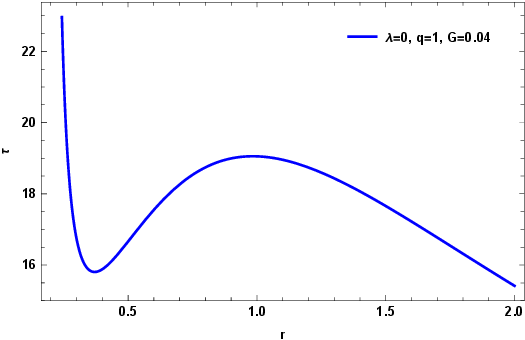}
 \label{1e}}
 \subfigure[]{
 \includegraphics[height=3cm,width=2.5cm]{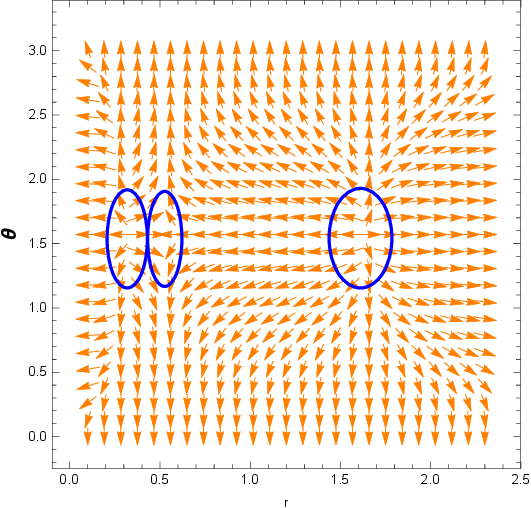}
 \label{1f}}
  \caption{\small{The curve described by Eq. (\ref{BBR3}) is illustrated in Figs. (\ref{1a}), (\ref{1c}), and (\ref{1e}). In Figs. (\ref{1b}), (\ref{1d}), and (\ref{1f}), the zero points (ZPs) are located at coordinates $(r, \theta)$ on the circular loops, corresponding to the free parameter ($\lambda, q, G$), $P=0.9$ and $\tau=6, 1, 17$}}
 \label{m1}
 \end{center}
 \end{figure}

Consequently, using Eqs. (\ref{N1}), (\ref{F1}), and (\ref{M7}), the Helmholtz free energy for this black hole is derived as follows,
\begin{equation}\label{BBR1}
\mathcal{F}=\frac{8 \pi  G P r^4+3 G q^2+3 r^2}{6 \pi  G P r}-\frac{\ln \left(\frac{\pi  \lambda  r^2}{G}+1\right)}{\lambda  \tau }
\end{equation}
Two vector fields, $\phi^{r_h}$ and $\phi^{\theta}$, are computed based on the previously discussed concepts and Eq. (\ref{F2}), as follows,
\begin{equation}\label{BBR2}
\phi^r=\frac{8 \pi  G P r^4-G q^2+r^2}{2 \pi  G P r^2}-\frac{2 \pi  r}{G \tau +\pi  \lambda  r^2 \tau }, \quad \phi^{\theta }=-\frac{\cot (\theta )}{\sin (\theta )}
\end{equation}
Additionally, we determine $\tau$ as follows,
\begin{equation}\label{BBR3}
\tau =\frac{4 \pi ^2 G P r^3}{\left(G+\pi  \lambda  r^2\right) \left(8 \pi  G P r^4-G q^2+r^2\right)}
\end{equation}
By analyzing these vector fields and the parameter $\tau$, we can gain valuable insights into the topological structure of phase transitions in black hole thermodynamics. This approach allows us to identify stable and unstable regions, as well as the nature of critical points, providing a deeper understanding of the underlying thermodynamic properties. In our study, we investigate the thermodynamic topology of AdS Reissner-Nordstr$\ddot{o}$m (R-N) black holes using nonextensive entropy, such as R$\acute{e}$nyi and Sharma-Mittal entropy, within two frameworks: bulk boundary and RPS thermodynamics. We first explore the thermodynamic topology in the bulk boundary framework. The illustrations are divided, with normalized field lines shown on the right. Figs. (\ref{m1}) and (\ref{m2}) display R$\acute{e}$nyi and Sharma-Mittal entropy, respectively. Figs. (\ref{1b}) and (\ref{1d}) reveal a singular zero point, indicating a topological charge determined by the free parameter $\lambda$ discussed in the study. This charge, which correlates with the winding number, is located within the blue contour loops at coordinates $(r, \theta)$. The sequence of these illustrations is governed by the parameter $\lambda$. The findings from these figures highlight a distinctive feature: a positive topological charge $(\omega = +1)$ and a total topological charge $W = +1$, represented by the zero point enclosed within the contour. Our analysis examines black hole stability by evaluating the winding numbers. Positive winding numbers suggest the thermodynamic stability of the on-shell black hole, further supported by specific heat capacity calculations. Given the solitary on-shell black hole, its topological number matches the winding number, amounting to 1. This indicates the presence of a single stable on-shell black hole, with a topological number reflecting a positive winding number across all black hole configurations $(W = \omega = +1)$.
\subsubsection{Thermodynamic topology within Sharma-Mittal statistics}
\begin{figure}[h!]
 \begin{center}
 \subfigure[]{
 \includegraphics[height=3cm,width=2.5cm]{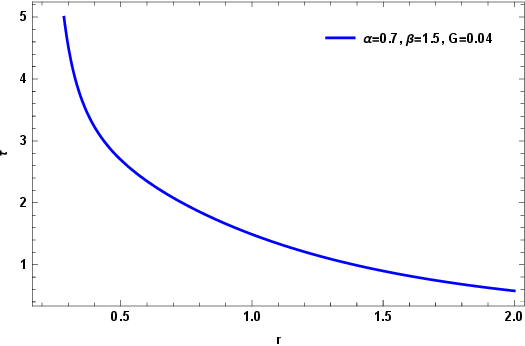}
 \label{2a}}
 \subfigure[]{
 \includegraphics[height=3cm,width=2.5cm]{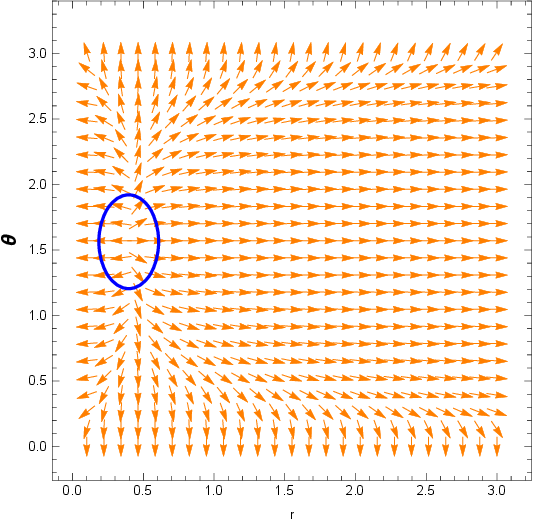}
 \label{2b}}
 \subfigure[]{
 \includegraphics[height=3cm,width=2.5cm]{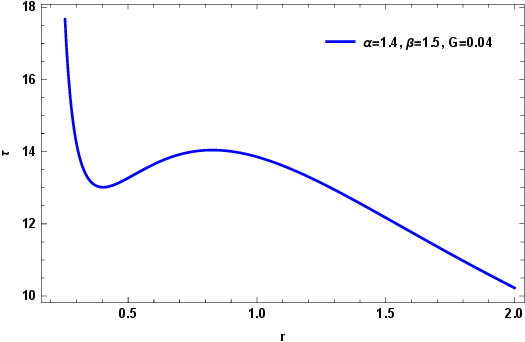}
 \label{2c}}
 \subfigure[]{
 \includegraphics[height=3cm,width=2.5cm]{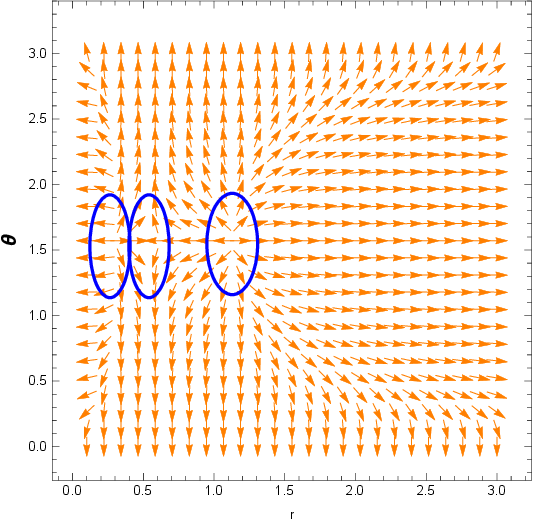}
 \label{2d}}
 \subfigure[]{
 \includegraphics[height=3cm,width=2.5cm]{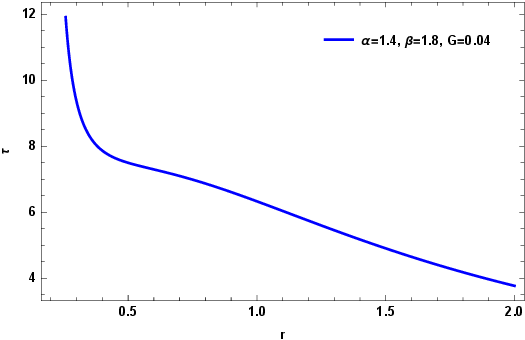}
 \label{2e}}
 \subfigure[]{
 \includegraphics[height=3cm,width=2.5cm]{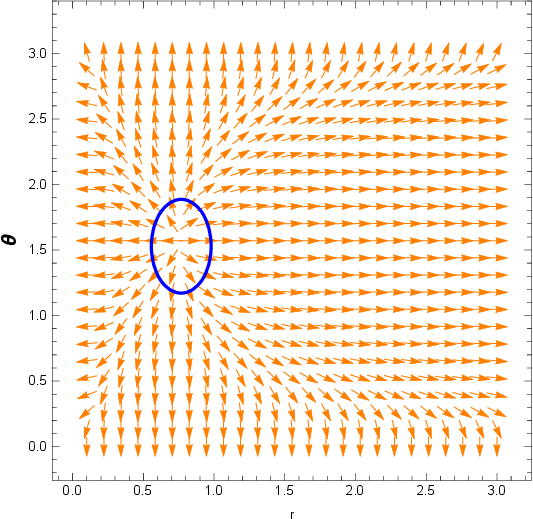}
 \label{2f}}
  \caption{\small{The curve described by Eq. (\ref{BBSM3}) is illustrated in Figs. (\ref{2a}), (\ref{2c}), and (\ref{2e}). In Figs. (\ref{2b}), (\ref{2d}), and (\ref{2f}), the zero points (ZPs) are located at coordinates $(r, \theta)$ on the circular loops, corresponding to the free parameter ($\alpha$, $\beta$ $G$), $P=0.9$ and $\tau=3, 13.5, 7$}}
 \label{m2}
 \end{center}
 \end{figure}

Here, utilizing Eqs. (\ref{N2}), (\ref{F1}), and (\ref{M7}), the Helmholtz free energy calculate as follows,
\begin{equation}\label{BBSM1}
\mathcal{F}=\frac{G q^2+r^2}{2 \pi  G P r}+\frac{-3 \left(\frac{\pi  \beta  r^2}{G}+1\right)^{\alpha /\beta }+4 \alpha  r^3 \tau +3}{3 \alpha  \tau }
\end{equation}
The $\phi^{r_h}$ is calculated based on Eq. (\ref{F2}), as follows,
\begin{equation}\label{BBSM2}
\phi^r=\frac{1}{2} \left(\frac{\frac{1}{\pi  P}-\frac{4 \pi  r \left(\frac{\pi  \beta  r^2}{G}+1\right)^{\frac{\alpha }{\beta }-1}}{\tau }}{G}-\frac{q^2}{\pi  P r^2}+8 r^2\right)
\end{equation}
Furthermore, $\tau$ is determined as follows,
\begin{equation}\label{BBSM3}
\tau =\frac{4 \pi ^2 G P r^3 \left(\frac{\pi  \beta  r^2}{G}+1\right)^{\alpha /\beta }}{\left(G+\pi  \beta  r^2\right) \left(8 \pi  G P r^4-G q^2+r^2\right)}
\end{equation}
Additionally, as shown in Fig. (\ref{1f}), when the parameter $(\lambda)$ is set to zero, our equations reduce to the Bekenstein-Hawking entropy structure, yielding different results. Fig. (\ref{1f}) shows the three topological charge $(\omega = +1, -1, +1)$ with total topological charge $W=+1$. Fig. (\ref{m2}) shows Sharma-Mittal entropy. As seen in Fig. (\ref{m2}), by increasing the parameter $\alpha$ from 0.7 to 1.4, the number of topological charges increases $(\omega = +1, -1, +1)$ with the total topological charge $(W = +1)$. Interestingly, when the parameter $\beta$ increases, the number of topological charges should decrease, resulting in $(\omega = +1)$ with the total topological charge $(W = +1)$, as clearly shown in Figs. (\ref{2b}), (\ref{2d}), and (\ref{2f}).
\subsection{RPS thermodynamics}
In the framework of restricted phase space (RPS), the entropy for an AdS R-N black hole is reformulated as,
\begin{equation}\label{M8}
q = \frac{\hat{q}}{\sqrt{C}},\quad G = \frac{l^2}{C},\quad S = \frac{C r_h^2 \pi}{l^2}
\end{equation}
The temperature $T$ of this black hole is:
\begin{equation}\label{M9}
T = \frac{C^2 l^2 r_h^2 + 3C^2 r_h^4 - l^4 \hat{q}^2}{4 \pi l^3 \sqrt{\frac{C^2r_h^2}{l^2}}Cr_h^2}
\end{equation}
The parameter $C$ is determined by,
\begin{equation}\label{M10}
C = \frac{3l^2 \hat{q}}{\sqrt{3l^2 - 9r_h^2} r_h}
\end{equation}
The critical value $r_c$ is, $\frac{\sqrt{6l}}{6}$. Using these equations, we can derive thermodynamic quantities such as mass as follows,
\begin{equation}\label{M10}
M = \frac{C^2 l^2 r_h^2 + C^2 r_h^4 + l^4 \hat{q}^2}{2 l^5 \sqrt{\frac{C^2 r_h^2}{l^2}}}.
\end{equation}
\subsubsection{Thermodynamic topology within R$\acute{e}$nyi statistics}
\begin{figure}[h!]
 \begin{center}
 \subfigure[]{
 \includegraphics[height=3cm,width=2.5cm]{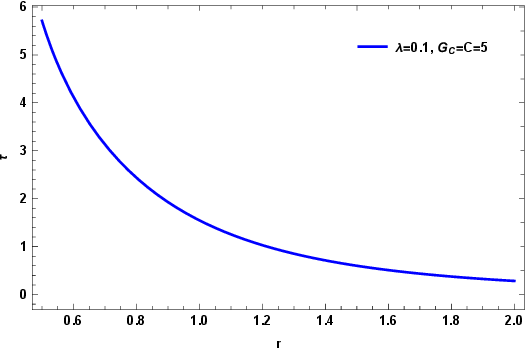}
 \label{3a}}
 \subfigure[]{
 \includegraphics[height=3cm,width=2.5cm]{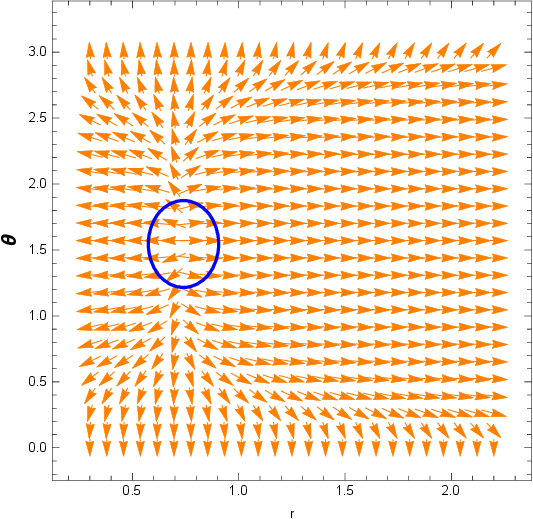}
 \label{3b}}
 \subfigure[]{
 \includegraphics[height=3cm,width=2.5cm]{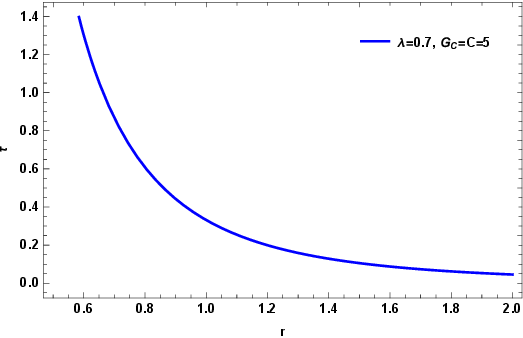}
 \label{3c}}
 \subfigure[]{
 \includegraphics[height=3cm,width=2.5cm]{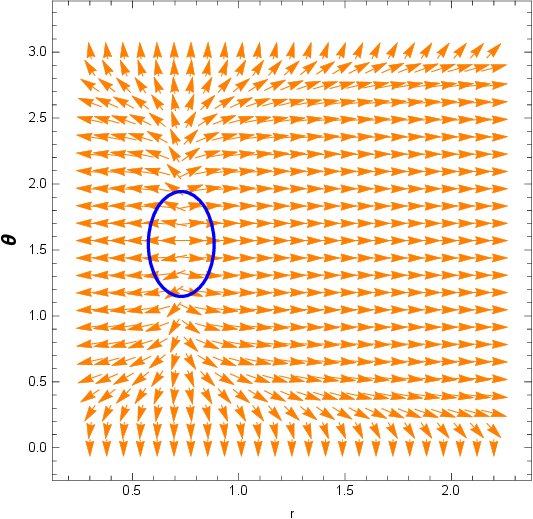}
 \label{3d}}
 \subfigure[]{
 \includegraphics[height=3cm,width=2.5cm]{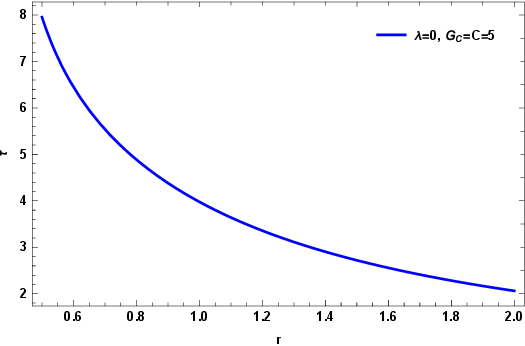}
 \label{3e}}
 \subfigure[]{
 \includegraphics[height=3cm,width=2.5cm]{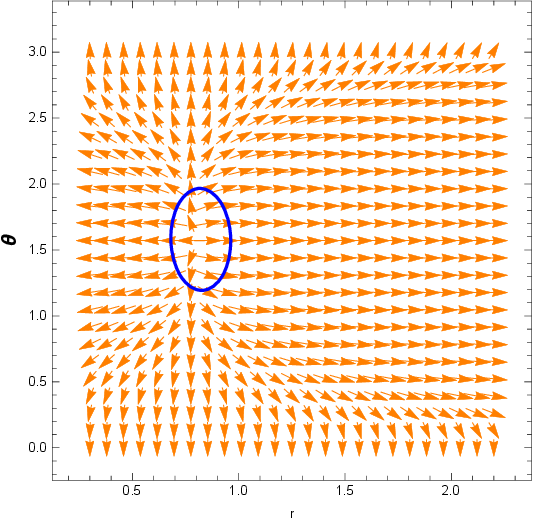}
 \label{3f}}
  \caption{\small{The curve described by Eq. (\ref{RPSR3}) is illustrated in Figs. (\ref{3a}), (\ref{3c}), and (\ref{3e}). In Figs. (\ref{3b}), (\ref{3d}), and (\ref{3f}), the zero points (ZPs) are located at coordinates $(r, \theta)$ on the circular loops, corresponding to the free parameter ($\lambda, C$), $q=1$, $l=1$ and $\tau=3, 0.8, 5$}}
 \label{m3}
 \end{center}
 \end{figure}

\begin{figure}[h!]
 \begin{center}
 \subfigure[]{
 \includegraphics[height=3cm,width=2.5cm]{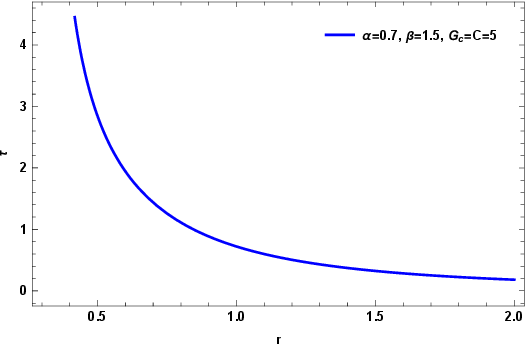}
 \label{4a}}
 \subfigure[]{
 \includegraphics[height=3cm,width=2.5cm]{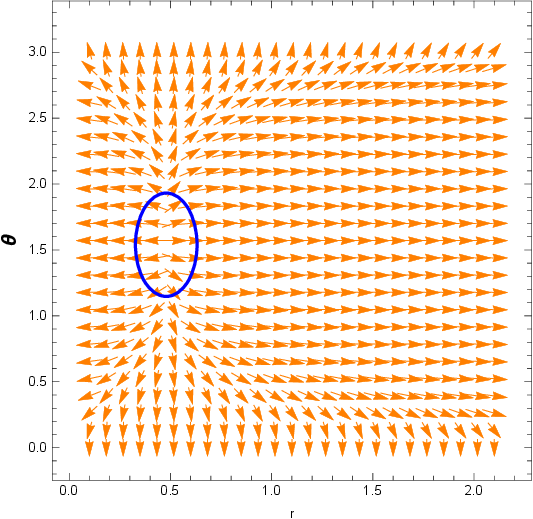}
 \label{4b}}
 \subfigure[]{
 \includegraphics[height=3cm,width=2.5cm]{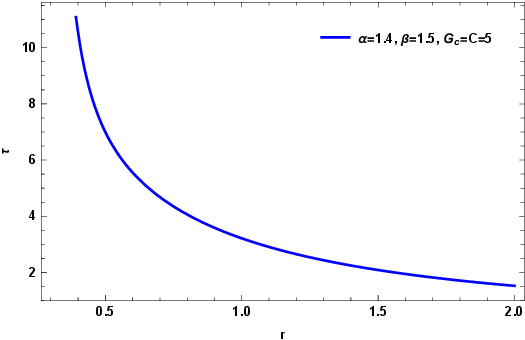}
 \label{4c}}
 \subfigure[]{
 \includegraphics[height=3cm,width=2.5cm]{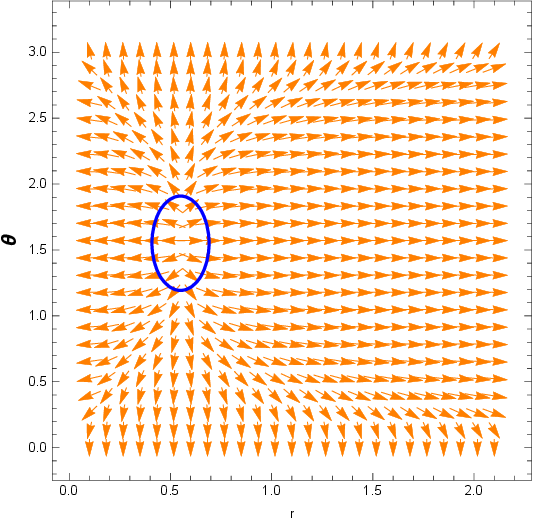}
 \label{4d}}
 \subfigure[]{
 \includegraphics[height=3cm,width=2.5cm]{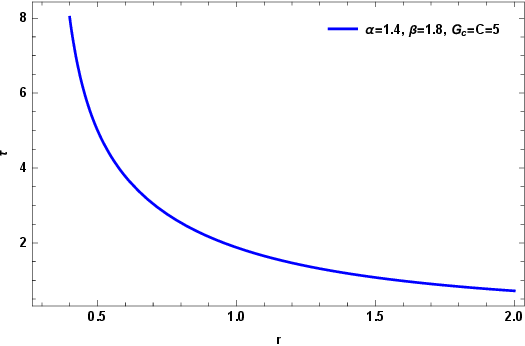}
 \label{4e}}
 \subfigure[]{
 \includegraphics[height=3cm,width=2.5cm]{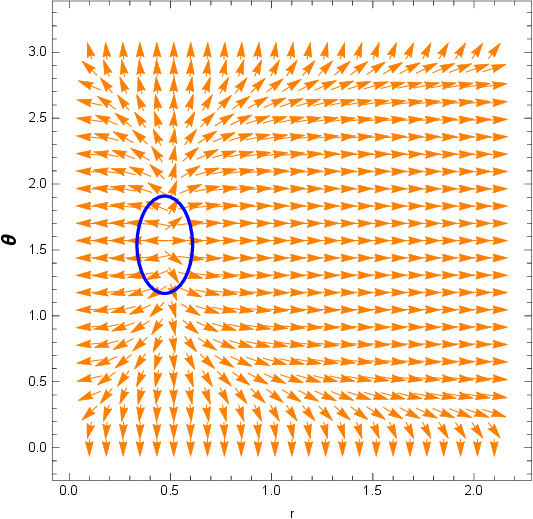}
 \label{4f}}
  \caption{\small{The curve described by Eq. (\ref{RPSSM3}) is illustrated in Figs. (\ref{4a}), (\ref{4c}), and (\ref{4e}). In Figs. (\ref{4b}), (\ref{4d}), and (\ref{4f}), the zero points (ZPs) are located at coordinates $(r, \theta)$ on the circular loops, corresponding to the free parameter ($\alpha$, $\beta$, $C$), $q=1$, $l=1$ and $\tau=3, 6, 5$}}
 \label{m4}
 \end{center}
 \end{figure}

For RPS thermodynamics, using Eqs. (\ref{N1}), (\ref{F1}), and (\ref{M10}), the $\mathcal{F}$ is derived as follows,
\begin{equation}\label{RPSR1}
\mathcal{F}=\frac{C^2 r^4+C l^2 r^2+l^4 \hat{q}^2}{2 l^5 \sqrt{\frac{C^2 r^2}{l^2}}}-\frac{\ln \left(\frac{\pi  C \lambda  r^2}{l}+1\right)}{\lambda  \tau }
\end{equation}
The $\phi^{r_h}$ is as follows,
\begin{equation}\label{RPSR2}
\phi^r=-\frac{l \hat{q}^2 \sqrt{\frac{C^2 r^2}{l^2}}}{2 C^2 r^3}+\frac{3 r \sqrt{\frac{C^2 r^2}{l^2}}}{2 l^3}+\frac{C r}{2 l^3 \sqrt{\frac{C^2 r^2}{l^2}}}-\frac{2 \pi  C r}{\pi  C \lambda  r^2 \tau +l \tau }
\end{equation}
Additionally, $\tau$ is determined as follows,
\begin{equation}\label{RPSR3}
\tau =\frac{4 \pi  C l^5 r^2 \sqrt{\frac{C^2 r^2}{l^2}}}{\left(3 C^2 r^4+C l^2 r^2-l^4 \hat{q}^2\right) \left(\pi  C \lambda  r^2+l\right)}
\end{equation}
A particularly interesting aspect of this study is its extension to the restricted phase space. When we continue our studies in this space with the two mentioned entropies, we notice that under all conditions and for all free parameters, we have a topological charge $\omega = +1$ with a total topological charge $(W = +1)$. Also, in RPS, when we reduce it to Bekenstein-Hawking entropy, similar behavior can be observed. This suggests that, unlike the bulk boundary space, similar behavior can be seen in both non-extended entropy and Hawking entropy states in RPS, as illustrated in Fig. (\ref{m3}) and Fig. (\ref{m4}).
\subsubsection{Thermodynamic topology within Sharma-Mittal statistics}
Here, with respect to Eqs. (\ref{N2}), (\ref{F1}), and (\ref{M10}), we will have,
\begin{equation}\label{RPSSM1}
\mathcal{F}=\frac{C^2 r^4+C l^2 r^2+l^4 \hat{q}^2}{2 l^5 \sqrt{\frac{C^2 r^2}{l^2}}}-\frac{\left(\frac{\pi  \beta  C r^2}{l}+1\right)^{\alpha /\beta }-1}{\alpha  \tau }
\end{equation}
The $\phi^{r_h}$ is calculated as follows,
\begin{equation}\label{RPSSM2}
\phi^r=\frac{C l^2 r^2 \left(\tau -4 \pi  l^2 \sqrt{\frac{C^2 r^2}{l^2}} \left(\frac{\pi  \beta  C r^2}{l}+1\right)^{\frac{\alpha }{\beta }-1}\right)+3 C^2 r^4 \tau -l^4 \hat{q}^2 \tau }{2 l^5 r \tau  \sqrt{\frac{C^2 r^2}{l^2}}}
\end{equation}
Additionally, $\tau$ is,
\begin{equation}\label{RPSSM3}
\tau =\frac{4 \pi  C l^5 r^2 \sqrt{\frac{C^2 r^2}{l^2}} \left(\frac{\pi  \beta  C r^2}{l}+1\right)^{\alpha /\beta }}{3 \pi  \beta  C^3 r^6+\pi  \beta  C^2 l^2 r^4+3 C^2 l r^4-\pi  \beta  C l^4 \hat{q}^2 r^2+C l^3 r^2-l^5 \hat{q}^2}
\end{equation}
In summary, our comprehensive analysis of the thermodynamic topology of AdS R-N black holes using nonextensive entropy frameworks reveals significant insights into the stability and phase transitions of these systems. The study underscores the importance of topological methods in understanding the complex thermodynamic behavior of black holes, paving the way for future research in this intriguing field.
\section{Conclusion}
In this paper, we explored the thermodynamic topology of AdS Reissner-Nordstr$\ddot{o}$m (R-N) black holes using nonextensive entropy frameworks, specifically R$\acute{e}$nyi (with nonextensive parameter $\lambda$) and Sharma-Mittal entropy (with nonextensive parameters $\alpha, \beta$). Our investigation spanned two frameworks: bulk boundary and restricted phase space (RPS) thermodynamics. In the bulk boundary framework, we encountered singular zero points that revealed topological charges influenced by the free parameter $\lambda$. These charges exhibited a positive topological charge $(\omega = +1)$ and the total topological charge $(W = +1)$, indicating the presence of a single stable on-shell black hole. Further analysis demonstrated that when $\lambda$ is set to zero, the equations align with the Bekenstein-Hawking entropy structure, showing different behaviors with multiple topological charges $(\omega = +1, -1, +1)$. Notably, increasing the parameter $\alpha$ in Sharma-Mittal entropy resulted in multiple topological charges $(\omega = +1, -1, +1)$ with a total topological charge $(W = +1)$. Conversely, increasing $\beta$ reduced the number of topological charges, maintaining the total topological charge $(W = +1)$. Extending our study to the restricted phase space, we observed consistent topological charges $(\omega = +1)$ across all conditions and parameters. This consistency persisted even when reducing to Bekenstein-Hawking entropy, suggesting similar behaviors in both non-extended and Hawking entropy states within RPS. Our comprehensive analysis underscores the significance of topological methods in understanding the complex thermodynamic behavior of black holes. These findings provide valuable insights into the stability, phase transitions, and black hole classification.
\section*{Authors' Contributions}
All authors have the same contribution.
\section*{Data Availability}
The manuscript has no associated data or the data will not be deposited.
\section*{Conflicts of Interest}
The authors declare that there is no conflict of interest.
\section*{Ethical Considerations}
The authors have diligently addressed ethical concerns, such as informed consent, plagiarism, data fabrication, misconduct, falsification, double publication, redundancy, submission, and other related matters.
\section*{Funding}
This research did not receive any grant from funding agencies in the public, commercial, or non-profit sectors.

\end{document}